\documentclass[twocolumn]{aastex631}

\newcommand\MBH{M_{\rm BH}}
\newcommand\Mhalo{M_{\rm halo}}
\shorttitle{Evolution of SMBH-halo mass relation}
\shortauthors{Aryan et al.}

\graphicspath{{./}{figures/}}

\usepackage{float}
\usepackage{amsmath}
\usepackage{mathtools}
\usepackage{fixltx2e}
\usepackage{ulem}
\usepackage{graphicx}
\usepackage{cancel}
\usepackage[caption=false]{subfig}

\begin{document}

\title{Evolution
of the mass relation between supermassive black holes and dark matter
halos across the cosmic time}

\author[0000-0002-6631-1886]{Aryan Bansal}
\affiliation{Department of Physics and Astronomy "Galileo Galilei"\\
University of Padova, Via Marzolo, 8 - 35131 Padova, PD, Italy}

\author{Kiyotomo Ichiki}
\affiliation{Graduate School of Science, Division of Particle and
Astrophysical Science, \\Nagoya University, Chikusa-Ku, Nagoya, Aichi, 464-8602, Japan}
\affiliation{Kobayashi-Maskawa Institute for the Origin of Particles and the
Universe, \\Nagoya University, Chikusa-ku, Nagoya, 464-8602, Japan}
\affiliation{Institute for Advanced Research, Nagoya University Furocho, Chikusa-ku, Nagoya, 464-8602, Japan}

\author{Hiroyuki Tashiro}
\affiliation{Graduate School of Science, Division of Particle and
Astrophysical Science, \\Nagoya University, Chikusa-Ku, Nagoya, Aichi, 464-8602, Japan}

\author[0000-0001-5063-0340]{Yoshiki Matsuoka}
\affiliation{Research Center for Space and Cosmic Evolution, Ehime University, Japan}



\begin{abstract}
A positive observational proof suggests that most galaxies contain a central supermassive black hole (SMBH) with mass in the range of $10^6M_\odot$ $-$ $10^{10}M_\odot$. It is suggested that the mass of SMBHs is proportionally related to that of the dark matter (DM) halo, even at $z= 6$. This implies that these SMBHs could coevolve with the host DM halos. Here, we investigate the mass evolution of SMBHs in a hierarchical structure formation by building halo merger trees using the extended Press-Schechter formalism. An SMBH with a mass that follows various power-law relations with the DM halo mass is assigned as an initial condition. Assuming that the mass growth of
all black holes is due to mergers, we obtain the relation between SMBH and DM halo masses at the present epoch. By requiring that the mass of the SMBHs at $z=0$ should not be greater than the currently observed SMBH-DM halo mass relation, a lower bound on the mass of the DM halo that can contain a SMBH can be imposed at $z= 6$ as $M_{\rm lim} > 3.6 \times 10^{10} M_\odot\times (1.4-n)^{2.3}$, where $n$ is the power-law index of the SMBH-DM halo mass relation at $z=6$. 
Because we only consider mergers for the mass evolution of SMBHs, this model is simplistic and should underestimate the mass of SMBHs relative to the mass of the DM halo at the present epoch. Our work aims to constrain the SMBH-DM mass relation at z=6 and not build a new model to explain the observations
\end{abstract}



\section{Introduction} \label{sec:intro}
According to observations, the majority of galaxies encompass a Super Massive Black Hole (SMBH). These black holes typically have masses of magnitude $10^{6} - 10^{10}$ $M_\odot$; for a review, see \citealt{2013ARA&A..51..511K}).
Many works attempt to explain the formation of black holes but the origin of SMBHs is still poorly understood. Some of the proposed theories include direct collapse (e.g., \cite{2003ApJ...596...34B,2020MNRAS.494.2851C}), collisions in dense stellar clusters (e.g., \cite{2020MNRAS.498.5652K,2002ApJ...576..899P,2021A&A...649A.160V}), mergers of stellar-mass black holes (e.g., \cite{2011ApJ...740L..42D,10.1093/mnras/staa2276}), Population III remnant model (e.g. \cite{2008MNRAS.385.1621K}, \cite{2019MNRAS.483.3592B}, \cite{2019MNRAS.487..198G}, \cite{2021MNRAS.501.4289Z}) and through supermassive stars or other progenitor objects (e.g., \cite{1972A&A....18...10A, 2021A&A...650A.204H}).
A solid and well-accepted proportionality relationship exists between the mass of these black holes and the mass of central bulge of the galaxies~\citep{2000ApJ...539L...9F, 2008IAUS..245..257S, 2015ApJ...815...21K, 2009ApJ...704.1135B, 10.1093/mnras/stz2472, 2019ApJ...877...64D}.  
Observations of QSOs have suggested the existence of SMBH at very high redshifts~\citep{2017ApJ...851L...8V,2018Natur.553..473B,2020ApJ...897L..14Y,2021ApJ...907L...1W}. 
Recent observations, including SDSS \citep{2001AJ....122.2833F}, CFHT \citep{2010AJ....139..906W}, Pan-STARRS \citep{2016ApJS..227...11B}, and HSC \citep{2019ApJ...883..183M,2022ApJS..259...18M}, have reported more than $200$ SMBHs shining as Quasi-Stellar Objects (QSOs) in the very early universe before the end of cosmic reionization era ($z \geq  6$). 
These SMBHs could have been caused solely by Inter and Intra-Galactic dynamics and stellar mechanics.
Many attempts have been made to define the evolution of SMBHs regarding galactic processes ~\citep{2000MNRAS.311..576K,2010A&ARv..18..279V,1998Natur.395A..14R, 2008mgm..conf..340D}, but none of the results have been proven.

Aside from the bulge-SMBH mass relationship, 
these SMBHs have shown strong correlations with the hosting DM halo, as inferred from \cite{2013ARA&A..51..511K}. A local relationship between mass of SMBH and that of DM halo has been well presented in \cite{2002ApJ...578...90F} where the author uses a correlation between rotational and central velocity dispersion in galaxies (originally discovered by \citealt{1979ApJ...234...68W}).
This halo-SMBH mass relationship is not only local but also at high redshifts.
Currently, for 49 QSOs in high redshifts,~$z \sim 6$, their velocity dispersion was measured and their DM halo mass was evaluated.
Based on these observations,
\citealt{2019ApJ...872L..29S} have reported a mass relationship even in high redshift ($z=6$). 

In this study, we show how a simple SMBH evolution with DM merger can connect the SMBH-DM halo relationship between $z=6$ and $z=0$.
We generate a merger tree based on the Press-Schechter formalism
and, while ignoring accretion, considered the growth of SMBH mass by merging.
By comparing the outcomes of the current study with the observations, we obtain a constraint on the nature of the dark matter halo with SMBHs, specifically, a lower limit on the mass of DM halo that can host the SMBHs.

Numerous works in the literature exist in which authors attempted to solve the challenges of galaxy evolution using various approaches.
\cite{2007MNRAS.382.1394M} investigated the hierarchical formation of galaxies using a semi-analytical approach described by \cite{2006RPPh...69.3101B}.
Their model suggested that for more massive SMBHs, the accretion is highly dominated  until $z=2$, but mergers begin to dominate in the mass development of SMBHs from $z=2$ to $z=0$.
Consequently, the mass growth total from $z=6$ to $z=0$ is dominated by mergers, and a small fraction of the final SMBH mass is from accretion.
\cite{2021RAA....21..212Z} also studied and compared the role of mergers and gas accretion in the evolution of SMBHs and host galaxies using a semi-analytical approach described by \cite{2013MNRAS.428.1351G}. 
 According to their findings, although the gas accretion dominates SMBH mass growth across all mass ranges, both mergers and QSOs mode accretion/starburst play a significant role in the evolution of SMBHs, which is backed up by the findings from \cite{2002MNRAS.335..965Y}.

The structure of this paper is as follows. The methodology is presented in the following section. We explain the Press-Schechter formalism used to generate our merger tree as well as the method used to fit the observations. We also describe how the SMBHs are assigned to progenitor halos.
Section 3 summarizes the results of the model concerning the evolution of simulated black holes from $z=6$ to $z=0$. In addition, the statistical presence of accretion in the formation of DM halos at $z=0$ as well as the role of merging/accretion in the evolution of SMBHs within these DM halos are also investigated. In section 4, we discuss different theoretical and practical aspects of the results. The outcomes of the current study, as well as the benefits and drawbacks of the model used in this study, are compared to other similar studies in the literature. Finally, in section 5,  we conclude our work. 

Throughout the paper, we adopt cosmological parameters: $h = 0.7010$, $\Omega_m = 0.287$, $\Omega_{\Lambda} = 0.692$ and $\Omega_m = 0.045$.

\section{Methodology}
\label{sec:methodology}
A merger history of DM halos satisfying the Press-Schechter mass function~(PSMF) from redshift $z=6$ to the present epoch was first constructed to investigate the connection between dark matter halos and SMBHs. 
An SMBH was assigned to each DM halo at $z=6$ as an initial condition for SMBH evolution,  using a power-law mass relationship between DM halos and SMBHs.
DM halo mergers were assumed to drive the mass evolution of SMBH and assess its mass evolution from redshift $z=6$ to redshift $z=0$. 
Finally, the obtained mass relationship between DM halos and SMBHs were compared with the observed one at the present epoch.

\subsection{Merger tree}
\label{sec:merger_tree}
A halo merger tree is a method that traces the historical formation of a dark matter halo by following the progression of its progenitor halos back in time.
Although several merger trees can be used to describe the evolution of DM halos (e.g., \citealt{1999MNRAS.305....1S}), we utilized the binary merger tree method \citep{1993MNRAS.262..627L}. 
This method is pretty easy to grasp while being very powerful to model the evolution of the DM halo merger.

The binary merger tree method assumes that each merger has two progenitors.
The merger rate is calculated using the extended PS formalism~\citep{1991ApJ...379..440B} which provides a DM mass function identical to that found by \cite{1974ApJ...187..425P} and also predicts the merger rate of DM halos.
At time $t$, consider a DM halo with mass $M$.
According to the extended PS formalism, the mass weighted probability that the mass of one progenitor is in the mass range between $M_{\rm p}$ and $M_{\rm p}+dM_{\rm p}$ at earlier time $t-\Delta t$, is given as
\begin{align}
   F(M_{\rm p}&,t-\Delta t|M, t) ~dM_{\rm p}
\nonumber \\
&= 
\frac{1}{\sqrt{2 \pi}} \frac{\Delta \delta}{\Delta S^{3/2}}
\exp \left[-\frac{(\Delta \delta)^2}{2 \Delta S } \right]
\left|\frac{dS(M_{\rm p})}{dM_{\rm p}} \right|dM_{\rm p}.
\label{eq:probability}
\end{align}
Here $\Delta S \equiv S(M_{\rm p}) - S(M)$ with the mass variance $S(M) = \sigma^2(M)$ and $\Delta \delta  \equiv \delta (t-\Delta t) - \delta (t)$ where $\delta(t) = \delta_c/D(t)$ with the critical linear density contrast $\delta_c=1.686$ for the spherical collapse~\citep{2010gfe..book.....M}, and the linger growth rate $D(t)$ at $t$.
In the matter dominated epoch, $D(t)$ is proportional to $1/(1+z)$.

To construct our merger tree, we start from the present epoch and move to high redshifts up to $z=6$. 
In the binary merger tree, at each time step $\Delta t$, one merger between two progenitors was assumed to occur.
The one progenitor mass, $M_{\rm p}$, is randomly chosen using the probabilities in Eq.~\eqref{eq:probability}, and the mass of the other progenitor is assigned to be $M-M_{\rm p}$. 
The procedure is then repeated for the obtained progenitors.

In constructing the merger tree, we also take into account accretion by introducing the lower limiting mass, $M_{\rm lim}$.
Progenitors were not considered if their masses were smaller than $M_{\rm lim}$.
Instead we assume that they accrete on the another progenitor as accreting diffuse dark matter and do not attempt further merger processes for them in subsequent time steps. In physical sense, limiting mass would correspond to a minimum mass above which halos can contain a SMBH.

For our merger tree, we use a constant timestep, $\Delta t$, which always corresponds to a $\Delta \omega \equiv \delta_c(\Delta z)=0.05$ (where, $\omega \equiv \delta_c(1+z)$), and  making around 235 steps of mergers or accretion before reaching redshift $z=6$.
We confirmed that our choice of the timestep makes sure that the DM halo abundance satisfies the PSMF at each timestep. 
For the limiting mass, we try different models in the merger tree and consider varying cases of $M_{\rm lim}=2\times 10^{6} M_\odot$, $7 \times 10^{6} M_\odot$, $2 \times 10^{7} M_\odot$, $7 \times 10^{7} M_\odot$, $2\times 10^{8} M_\odot$, $7 \times 10^{8} M_\odot$, $2 \times 10^{9} M_\odot$.

\begin{figure}[h!]
\centering
\includegraphics[width=9cm,height=10cm]{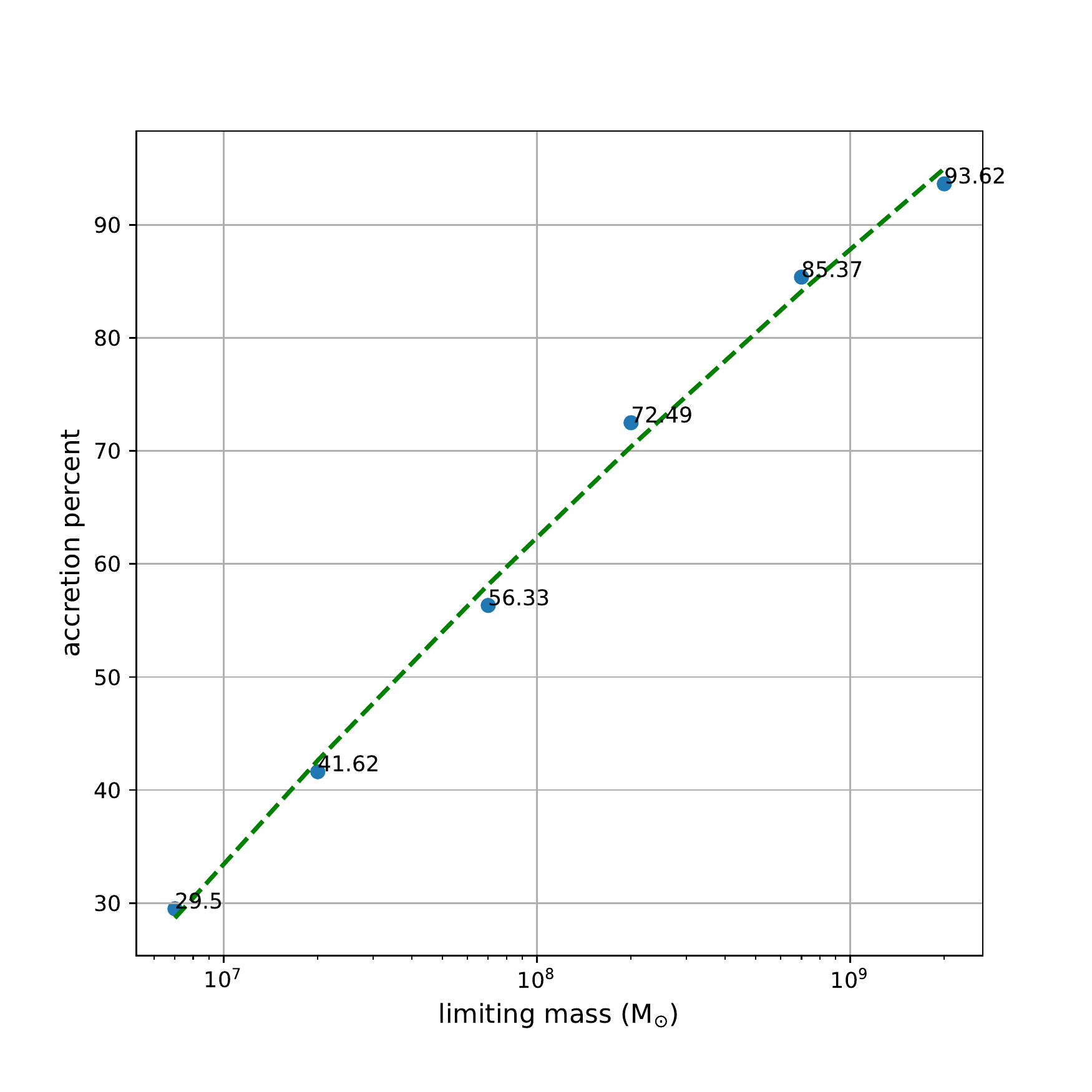}
\caption{The accretion rate vs the limiting mass($M_{\odot}$). The green-dotted line represents the interpolated curve which is a logarithmic fit.}
\label{fig:accrrate}
\end{figure}

The DM halo mass evolves via the merger of progenitors and through accretion.
The ratio of mass due to the merger and the accretion is determined by the limiting mass.
Figure~\ref{fig:accrrate} depicts the accretion ratio in the mass evolution of dark matter halo as a function of $M_{\rm lim}$.
The large limiting mass significantly increases the contribution of accretion in the DM halo mass growth.

\subsection{SMBH evolution}
Our model assumes that the SMBH mass can grow through the merger of SMBHs residing in the merging progenitor DM halos.
As a starting point of our simulation, we assign an SMBH to each DM halo with a mass larger than $M_{\rm lim}$ at $z=6$ that is created in the merger tree described in Sec.~\ref{sec:merger_tree}.

\begin{figure}[h!]
\centering
\includegraphics[width=8cm,height=10cm]{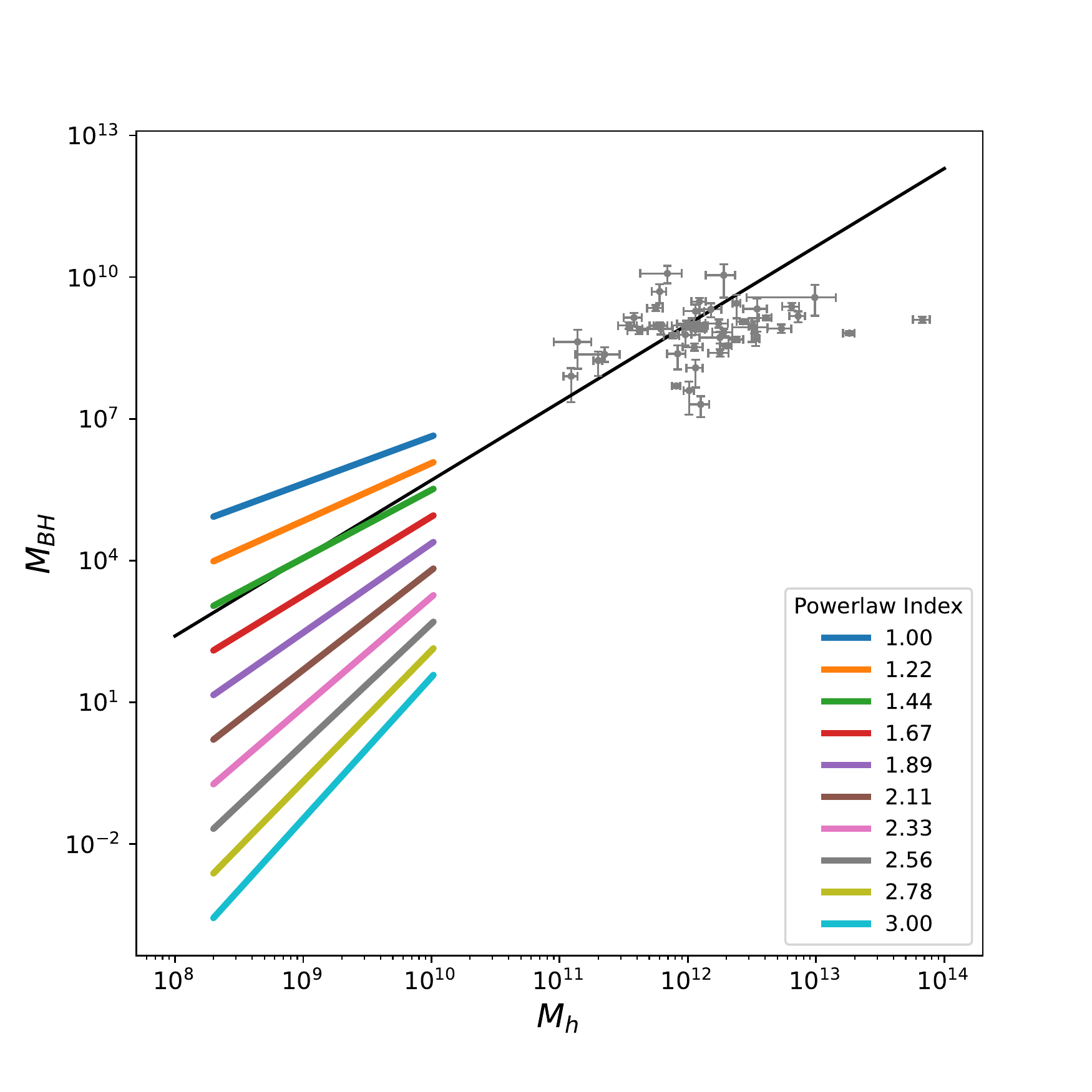}
\caption{SMBH-Halo relationship at $z=6$. The thin solid line with a slope fixed at $n=1.65$ is the best fitting line for the observation data. The other lines represent the power-law relationships used in the analysis defined in Eq.~(\ref{eq:initial_powerlaw}), with the slopes indicated by the color coordinates.}
\label{fig:z6multirel}
\end{figure}

\cite{2019ApJ...872L..29S} investigated the relationship between the SMBH and the DM halo mass in 49 QSOs 
at $z \sim 6$.
Motivated by this work, we assign the mass of SMBHs following a simple power-law mass relationship between SMBHs and DM halos,
\begin{equation}
\frac{M_{\rm BH}}{M_0}=A\left(\frac{M_h}{M_1}\right)^n~,
\label{eq:initial_powerlaw}
\end{equation}
where $M_1, M_0$ is the pivot mass of the relationship given as $(M_1,M_0)=(3.5\times 10^{12} M_\odot,1.5\times 10^9 M_\odot)$. 
Changing the amplitude $A$ in the equation at the initial condition simply changes the amplitude in the outcomes proportionally. To determine this value, we used the observed $z=6$ black hole mass function from \cite{2010AJ....140..546W} and abundance matched with our halo mass function. The results of the abundance matching shows that the value of A lies in a range of 1 to 3 where A is larger for smaller values of powerlaw index n.
To find the constraints on the mass relation, we fixed $A$ to be unity in the following analysis.
To examine the impact of the initial condition on the mass relationship between SMBHs and DM halos at $z=0$, we take the different power-law indices $n$ in $1.0<n<3.0$. 
In Fig.~\ref{fig:z6multirel}, the colored lines represent our initial power-law relations.
The data points in the figure represent 49 QSOs investigated in~\cite{2019ApJ...872L..29S}.

The observations at high redshifts ($z\sim 6$) tend to favor more massive objects. 
The observed DM halo mass is larger than the typical mass in our merger trees.
However, the observations may have missed numerous less-massive DM halos which also have SMBHs with smaller masses.
Thus, we selectively adopt the initial conditions which include SMBH with a smaller mass than indicated by the observed relation.

In our model, the merger of two DM halos following the merger tree is accompanied by the merging of the SMBH hosted by the merging DM halos.
We assume that, during the merging, the SMBH mass is conserved similar to the DM halo mass in the merger tree and we ignore the SMBH mass growth due to the accretion. Hence the total SMBH mass is conserved. Later, we also consider cases where mass is lost through gravitational waves following the mergers and study how the mass-loss from gravitational waves affect the evolution of SMBHs.

\section{Results}
\label{sec:Results}
\subsection{$\MBH$-$\Mhalo$ relation at $z=0$}

\begin{figure*}[h!]
\centering
\includegraphics[height=6cm]{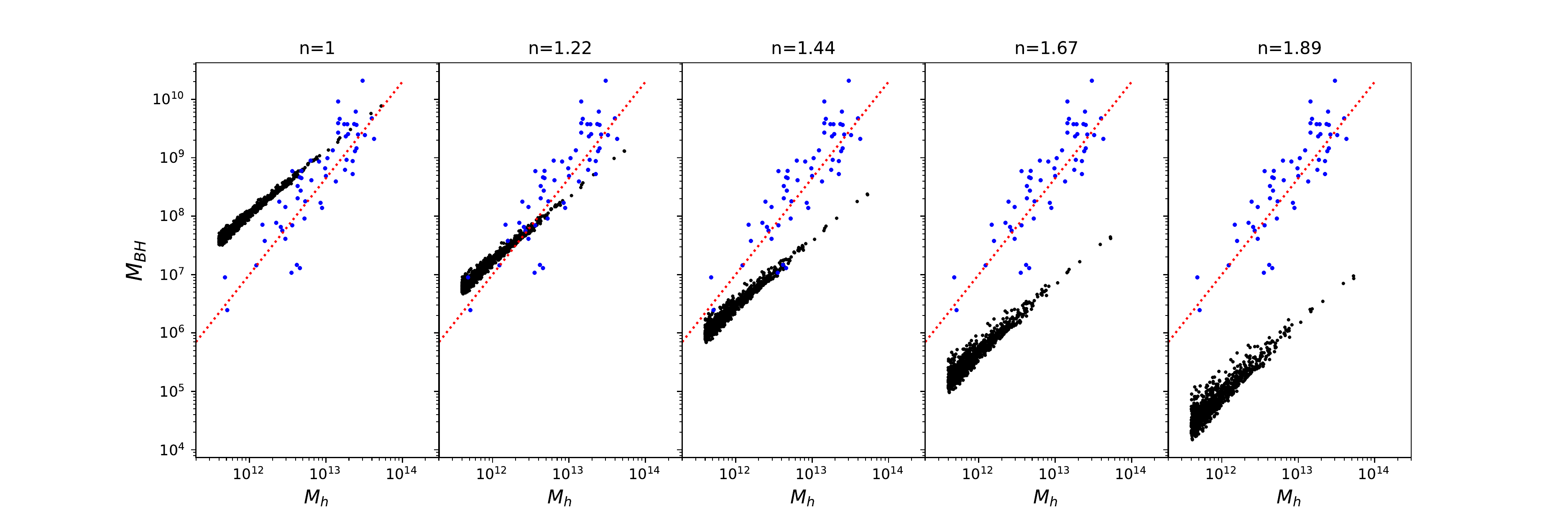}   
\caption{Final mass distribution of SMBHs in our simulation (black points) with a limiting mass of $2\times 10^8 M_{\odot}$ for five different initial power-law indices (at $z=6$) as indicated on top of each graph, compared with the observations at z=0 (blue points). The red dotted line is the local relation by \cite{2002ApJ...578...90F}}.
\label{fig:bh02e8}
\end{figure*}

In the model presented in this study, SMBHs evolve via several mergers from $z=6$ to $z=0$.
During this evolution, the mass relationship between SMBHs and DM halos changes. 
Thus, we plot the resultant mass relationships between SMBHs and DM halos at $z=0$ for five different initial power-law indices in  Fig.~\ref{fig:bh02e8}.
The limiting mass is set to $M_{\rm lim} = 2\times 10^{8}M_\odot$ in this case. 
For comparison, 85 observed SMBH data points taken from \cite{2002ApJ...578...90F} are plotted. They investigated the mass relationship at $z=0$ and the resulting relationship is depicted as the red dotted line in  Fig.~\ref{fig:bh02e8}.

Figure~\ref{fig:bh02e8} shows that, as $n$ decreases, the mass of SMBHs at $z=0$ increases for all DM halos of different masses.
When $n$ is small, SMBHs in small DM halos have greater masses than when $n$ is large, and the total SMBH mass increases. Without mass loss, the total SMBH mass is conserved in our model, and SMBHs at $z=0$ have greater mass as well, and the mass relationship is amplified for small $n$.
However, the slope of the mass relationship at $z=0$ does not strongly depend on the initial power-law index~$n$.
We will discuss this point later.

\begin{figure*}[h!]
\centering
\includegraphics[height=9cm,keepaspectratio]{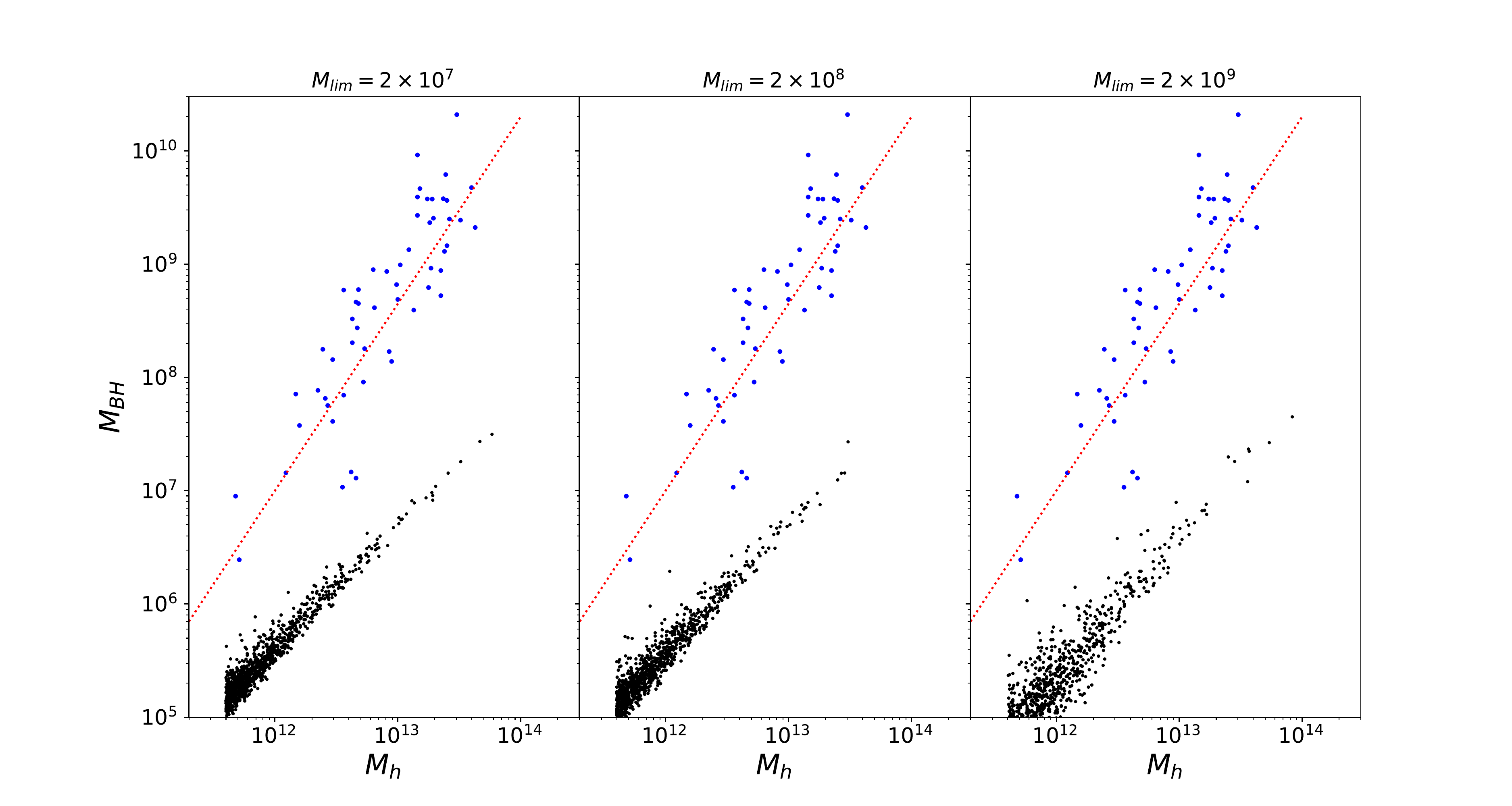}   
\caption{Comparison of the final mass distribution at $z=0$ for three different limiting masses, $2\times 10^7M_\odot$, $2\times 10^8M_\odot$ and $2\times 10^9 M_{\odot}$ from left to right. The power-law index for each case has been fixed to $n=1.67$. Again, the black points are the black holes in our model, blue are the observed black holes and the red-dotted line is the local relation}
\label{fig:bh0compare}
\end{figure*}

Figure ~\ref{fig:bh0compare} represents the effect of the limiting mass on the mass relationship at $z=0$ with a fixed initial power-law index $n=1.67$.
In the figure we take 
$M_{\rm lim} = 2\times 10^7M_\odot$, $2\times 10^8M_\odot$ and $2\times 10^9 M_{\odot}$ from left to right. As the limiting mass increases, the number of SMBH hosting halos, and hence the number of SMBHs, at z=6 decreases. Hence, the total mass of the SMBHs at z=0 decreases. This decrease in SMBH mass at $z=0$ is not as significant when compared to the decrease in mass due to increasing powerlaw index at $z=6$.

\begin{figure}[h!]
\includegraphics[width=8cm,height=7cm]{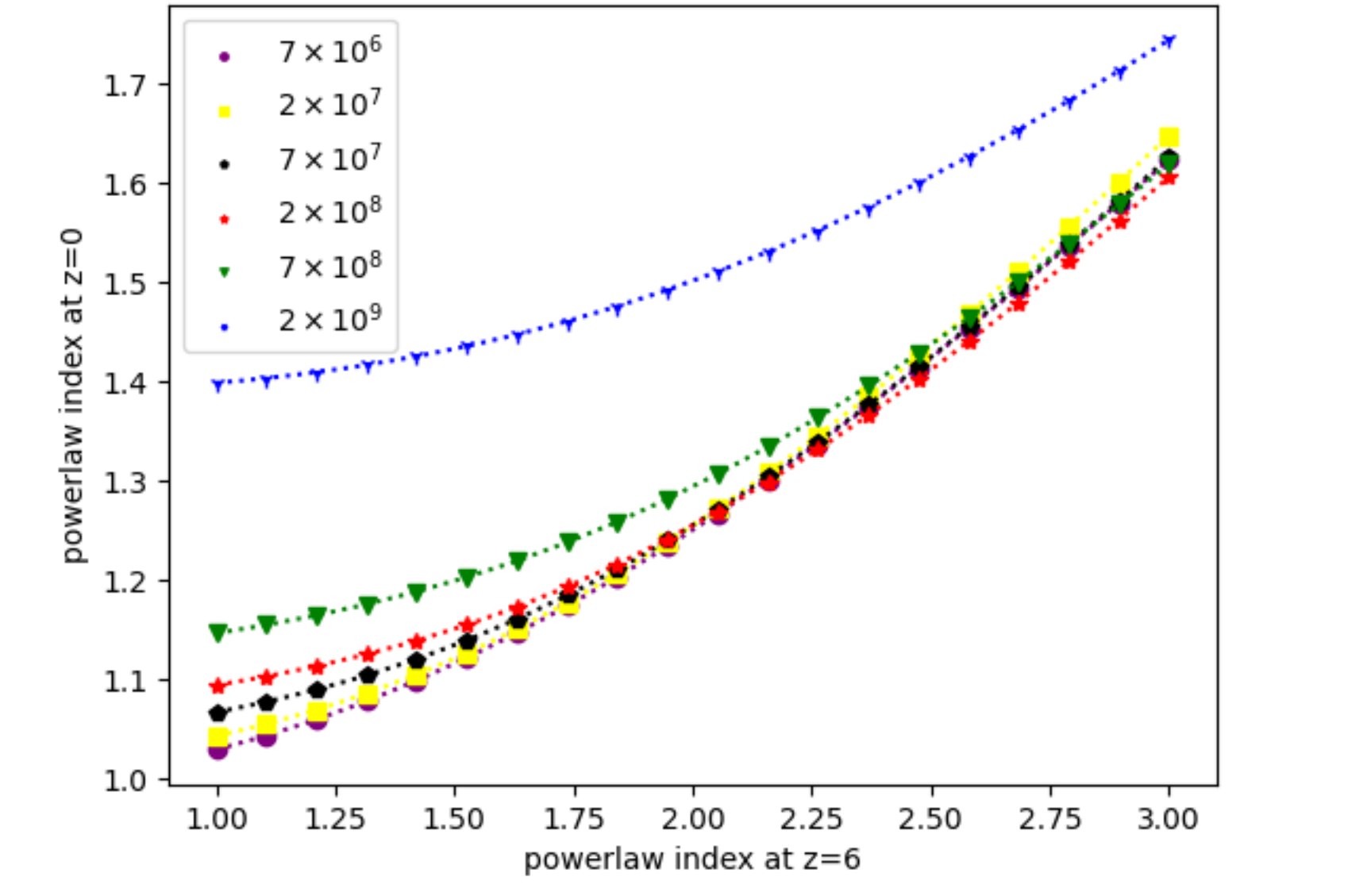}   
\caption{Relation between power law indices at $z=6$ and $z=0$ for different limiting mass (in $M_{\odot}$).}
\label{fig:powevol}
\end{figure}

\begin{figure}[htb!]
\includegraphics[width=8cm,height=8cm]{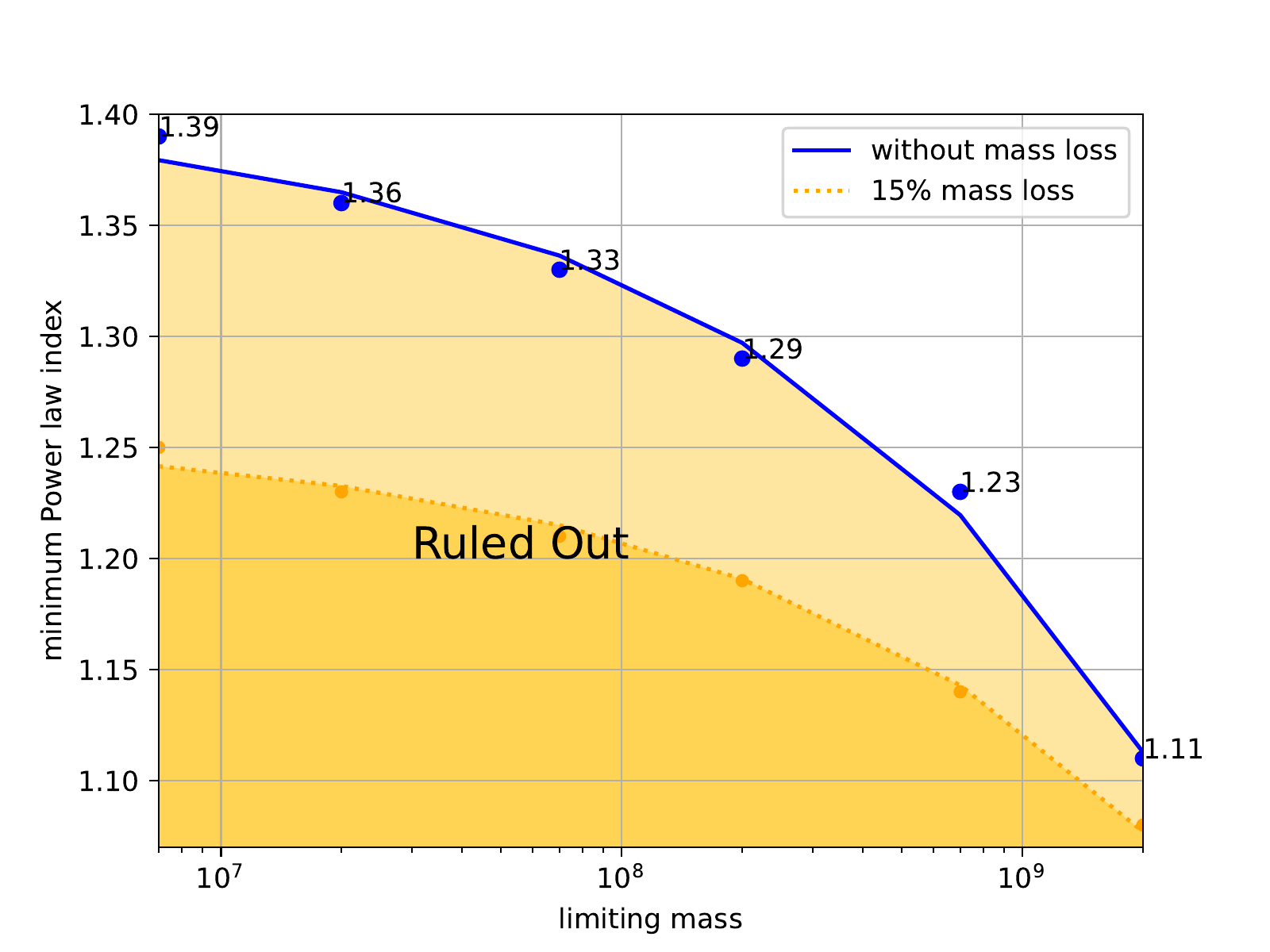}   
\caption{Constraints on the power-law index and the limiting mass required at $z=6$ to explain the observations at $z=0$, fitted with powerlaw functions. The filled region is excluded from the DM halo-SMBH mass relationship at $z=0$. The blue points are the constraints in the case of no mass loss due to gravitational waves. The orange points are in the model where each merger has a 15\% mass lost due to gravitational waves.}
\label{fig:minindex}
\end{figure}

To examine the dependence of the slope in the mass relation at $z=0$, we fit the mass relationship between SMBHs and DM halos at $z=0$ with a single power-law function to obtain the best-fit power-law index.
The results are plotted  in Fig.~\ref{fig:powevol} as a function of the initial power-law index.
The different colored lines in the figure represent the various limiting mass conditions. 
The power-law index at $z=0$ is directly correlated with the index at $z=6$. 
As the initial power-law index increase, so does the resultant index at $z=0$. 

Figure~\ref{fig:powevol} also shows the dependence of the mass relationship on the limiting mass at $z=0$.

\subsection{Constraints on the limiting mass and the slope of the $M_{\rm BH}$-$M_{\rm halo}$ relationship}

We have ignored the accretion growth of the SMBH mass in our model even though the accretion may contribute significantly to the SMBH mass growth.
Consequently, the resulting SMBH mass of the current study hosted by DM halos with mass $M_{\rm h}$ may be the lower bound of the SMBH mass hosted by DM halos with mass $M_{\rm h}$ for the initial condition set at $z=6$.
Thus, when the SMBH mass in our simulation is greater than the mass relation observed at $z=0$ which is represented as red lines in Figs.~\ref{fig:bh02e8} and \ref{fig:bh0compare}, the initial condition, the initial power-law index $n$, and the limiting mass $M_{\rm lim}$, can be ruled out because the observation does not favor such massive SMBH for DM halos with $M_{\rm h}$.

Figure~\ref{fig:minindex} depicts the resulting constraint.
The ruled-out parameter region is represented in the figure by the yellow shaded region.
Accordingly we obtain the fitting formula of $(M_{\rm lim},n)$ shown with the blue line in the figure as
\begin{equation}
M_{\rm lim} > 3.6 \times 10^{10} M_\odot\times (1.4-n)^{2.3}~.    
\end{equation}
The initial parameter set $(M_{\rm lim},n)$ must satisfy the above equation, at least, not to violate the observed mass relationship between SMBHs and DM halos at $z=0$.

Finally, a slight difference between the slopes of our model (for example, $n\approx 1.2$ for the initial power-law index $1.75$ at $z=6$) and observations ($n=1.65$) at $z=0$ could be explained by gas accretion onto BHs, implying that the more massive black holes require more accreting gas than less massive ones to satisfy the observed slope. Another explanation could be a different form of mass relation at $z=6$ compared to Equation (\ref{eq:initial_powerlaw}).

\subsection{Mass loss due to gravitational waves}

Black hole mergers are cataclysmic events that produces strong gravitational waves. General relativity predicts that gravitational waves emitted by the shrinking binary during the final coalescence carry away a non-zero net linear momentum, so that the centre of mass of the merged black holes recoils (Peres 1962; Campanelli et al. 2007; Miller et al. 2007). Along with momentum, gravitational waves also carry away energy. This results in a net loss of the black hole mass. Hence, the final black hole born by merger has a smaller mass than the sum of the masses of the
two progenitors. \cite{2010MNRAS.401.2021R}

A constraint on the mass of the black holes can be provided such that the area of the event horizon does not decrease. This condition, thus, translates to: $M_{remnant}^2 >  M_1^2 + M_2^2$
where $M_1$ and $M_2$ are the masses of two progenitor black holes\\
If the two black holes are equally massive, the above equation becomes:
$M_{remnant} >  \sqrt2 \times M_1$

This scenario results in a mass loss of $\frac{2-\sqrt2}{2}$, i.e. 29\% of the mass potentially radiated away.
Since, this is a cumulative effect, and even though 29\% is an extreme example of both black holes of same mass, even small mass loss rates can conceivably cause significant mass loss.

\begin{figure}[!htp]

\subfloat[]{
  \includegraphics[clip,width=\columnwidth]{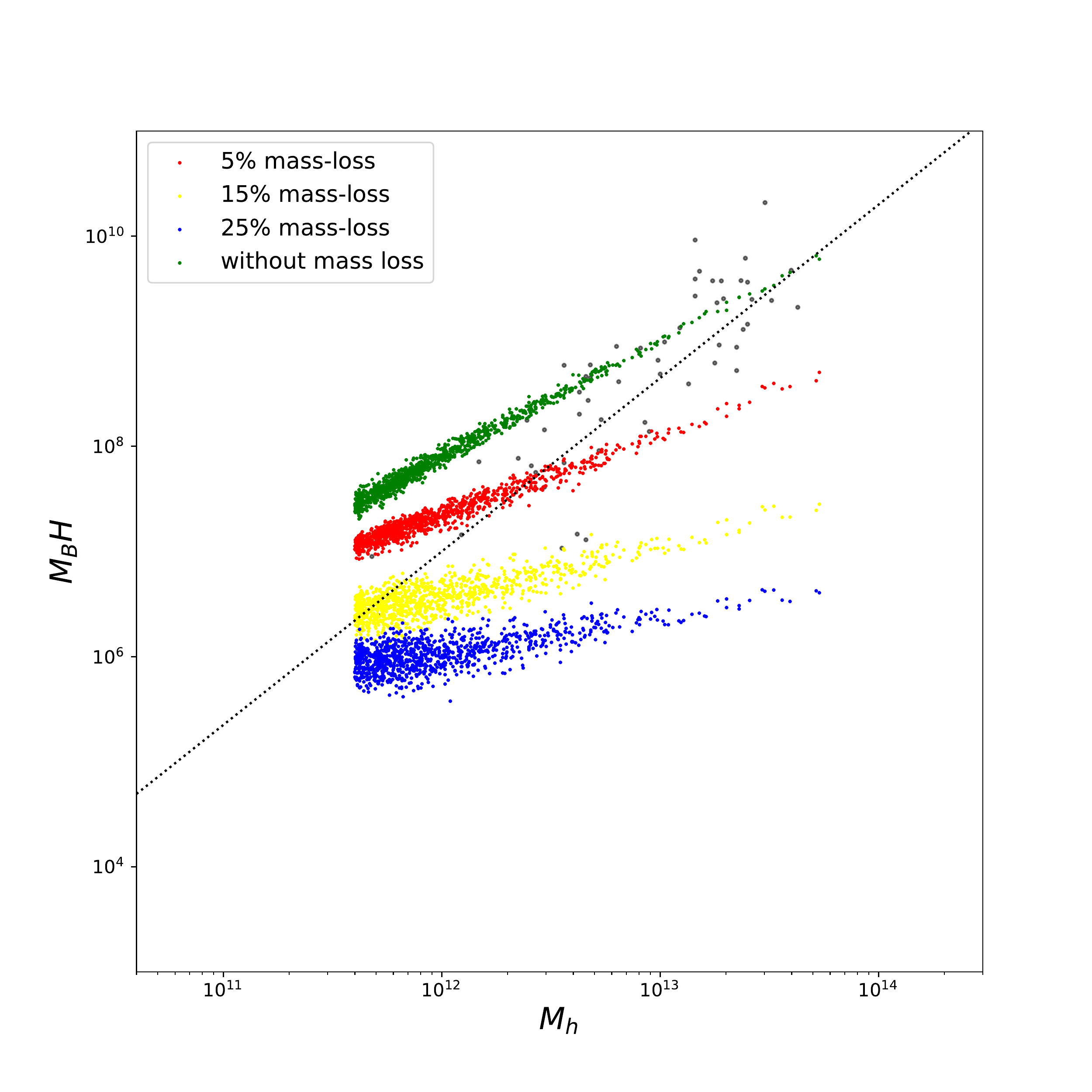}
}

\subfloat[]{
  \includegraphics[clip,width=\columnwidth]{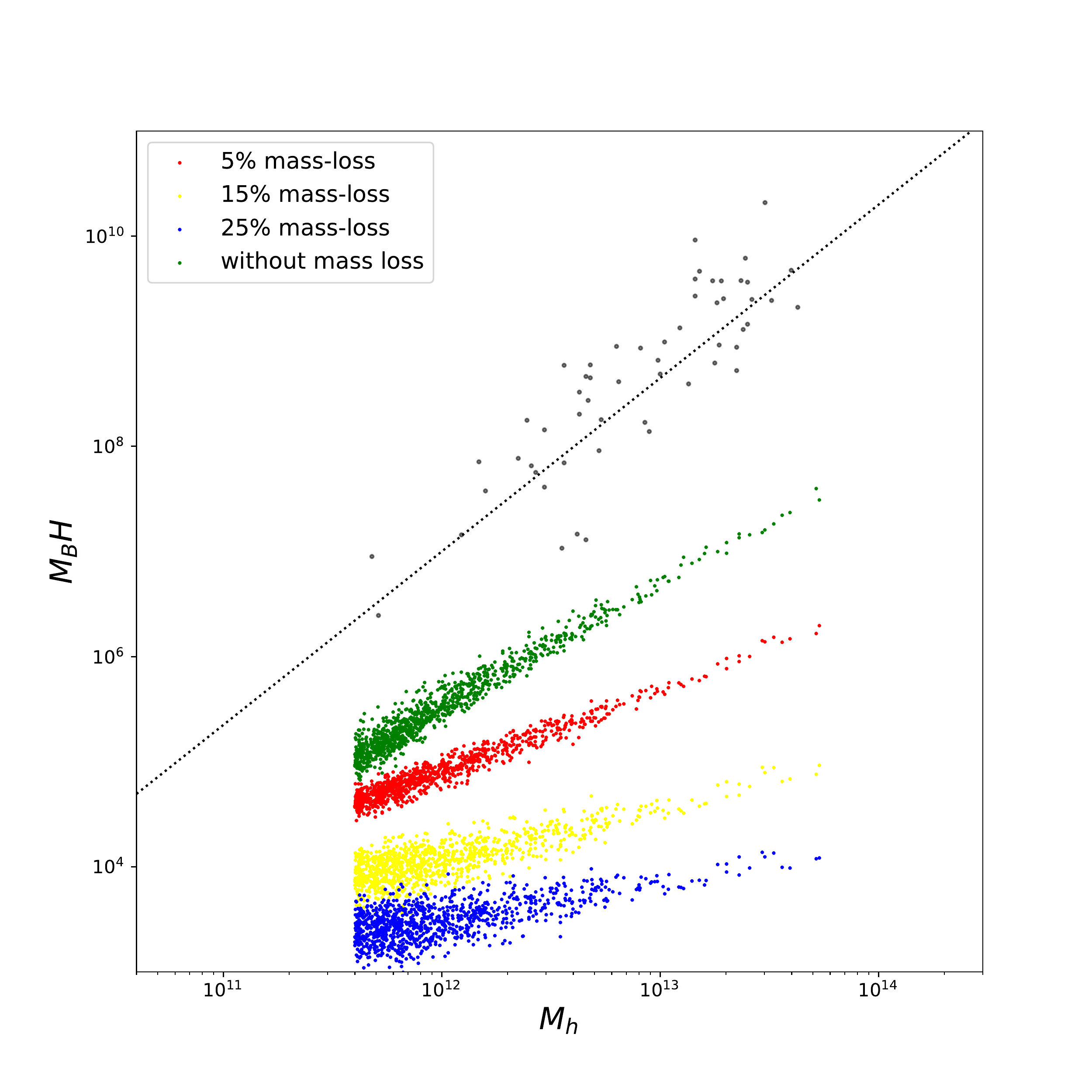}
}

\caption{Effect of mass-loss on the evolution of black holes: (a) and (b) are for cases with initial powerlaw index of SMBH-halo mass relation of n=1 and n=1.65 respectively}
\label{fig:massloss}
\end{figure}

To approach this concern and understand the affect of mass loss on the evolution of black holes, we introduced a mass loss in our black hole merger history. We considered three scenarios where each black hole merger is affected by a 5\%, 15\% and a 25\% mass-loss respectively. The results have been plotted in Figure \ref{fig:massloss} for a limiting mass of $2 \times 10^8 M_\odot$. In our simulations, a 5\% mass-loss in each merger causes on average, an overall 80\% loss in the final mass of the black hole. We also notice a 95\% and 99.2\% lost mass on average in total for the 15\% and 25\% case respectively. Hence, a considerable amount of mass is lost from gravitational waves during the merger history. The result of this massloss on the initial power law index constraint of the mass relation is shown in Figure \ref{fig:minindex}. For the case of 15\% mass loss, this constraint, as shown by the orange dotted line in the figure, is best fitted by the formula:
\begin{equation}
    M_{\rm lim} > 1.2 \times 10^{12} M_\odot\times (1.3-n)^{4}~
\end{equation}

\section{Summary and Discussion}
\label{sec:Summary}
The goal of this study is to better understand the impact of mergers and find the minimum mass of halos required to explain the observations and thus host an SMBH. 
To connect the SMBH-DM halo mass relations at $z=0$ and $z=6$, a simple merger tree model was used. 
Our simple model was unable to account for the gas accretion onto BHs. 
Along with mergers, gas accretion onto black holes plays a vital role in nature.
In particular, gas accretion could make a significant contribution to the mass growth of SMBHs.
Thus, the final SMBH massses obtained in this work are at a lower bound 
for a given initial SMBH mass distribution at $z=6$.

As a part of our merger tree simulation, we introduced a redshift independent lower limit in the halo mass, below which the dark matter was treated as accretion.  
In other words, halos were assumed to not exist below this limiting mass or that they did not contain a central Black Hole.
We found that the slope of the final relationship between $M_h$ and $\MBH$ at $z=0$ was nearly independent of this limiting mass, while both the initial slope and the limiting mass change the amplitude of the final black hole mass.

This can be obvious by regarding a dark matter halo with mass $m_{\rm halo}$ at $z=0$ w
hich is formed by mergers of ${d N(M,m_{\rm halo})}/{dM} \Delta M$ dark matter halos at $z=6$ in the mass range between $M$ and $M+\Delta M$,
\begin{align}
m_{\rm halo} = \int d M  M\frac{d N(M,m_{\rm halo})}{dM}.
\end{align}
Let this dark matter halo with mass $m_{\rm halo}$ at $z=0$ have a massive black hole with mass $m_{\rm BH}$ given by
\begin{align}
m_{\rm BH} = \int_{M_{\rm lim}} d M  M_{\rm BH}(M) \frac{d N(M,m_{\rm halo})}{dM},
\end{align}
where ${M_{\rm lim}} $ is the minimum mass of dark matter halo for hosting a BH at $z=6$ and $M_{\rm BH}(M) $ is the $M_{\rm bh}-M_{\rm halo}$ relationship at $z=6$.

If $d N(M,m_{\rm halo})/dM$ depends only on $m_{\rm halo}$ and not on $M$, i.e., ${d n(M,m_{\rm halo})}/dM = f(m_{\rm halo})$
we obtain
\begin{align}
m_{\rm BH} =& \int_{M_{\rm lim}} d M  M_{\rm BH}(M) \frac{d N(M,m_{\rm halo})}{dM}
\nonumber \\
=& f(m_{\rm halo}) \int_{M_{\rm lim}} d M  M_{\rm BH}(M) 
\nonumber \\
=& f(m_{\rm halo}) A(M_{\rm lim})~.
\label{eq:universal_index}
\end{align}
Therefore, the power-law index of the $m_{\rm BH}-m_{\rm halo}$ relation at  $z=0$ is determined by the function $f(m_{\rm halo})$, the Press-Schechter mass function, if we ignore gas accretion onto the BHs.

The power-law index at $z=0$ does not change for various $M_{\rm lim}$  in cases with less limiting mass~($M_{\rm lim} \lesssim 7 \times 10^7 M_\odot$) as can be seen in figure \ref{fig:powevol}.
However, in cases with a large limiting mass~($M_{\rm lim}\gtrsim 7 \times 10^7 M_\odot$),
a larger limiting mass results in a larger power-law index at $z=0$. 
This reason is mainly due to the shape of the halo mass function. 
 For a simple power-law halo mass function, a fixed index can be seen from Eq.~\ref{eq:universal_index}. 
Since our halo mass function follows the PSMF, the exponential tail can significantly affect the slope of the SMBH-halo mass relationship at $z=0$.

In figure \ref{fig:powevol}, we plotted the constraints on the initial powerlaw index and the limiting mass. A significant amount of mass is lost during the black hole mergers in the form of gravitational waves which causes a relaxation in this constraint, thus giving more freedom of choice in the initial power law index of the mass relation. The mass loss causes the total mass of black holes shown in figure \ref{fig:bh02e8} to decrease, thus having more scenarios where the total SMBH mass is lower than the observed mass and allowing more smaller values of initial powerlaw index to be valid.

In our analysis, we also considered the effect of initial scatters on the final results. The presented results are for cases without initial scatter. Introducing an initial scatter partially propagates to the final one, and the final scatter becomes lesser after encountering multiple mergers. For example, when Gaussian scatter around the initial power-law relationship is considered with a variance of about two orders of magnitude, the variance in the final relation is less than half the order of magnitude. The variance of the observed relationship (blue points) is about one order of magnitude. Thus, one needs an extra source of the scatter, most likely from gas accretion onto the SMBHs to explain this observed scatter.

\cite{2021MNRAS.507.4274M} investigated the mass relationship between SMBH and the host galaxy using a simple equilibrium model.
Their work was heavily influenced by \cite{10.1093/mnras/stw2735} who primarily assumed equilibrium between cosmological accretion onto halos, gas radiative cooling, star formation, BH accretion, and simplified treatments for stellar and AGN feedback to model their galactic setting. 
They proposed a physical scenario to explain the observed mass relationship, but their results cannot be directly compared with the current study because they ignore the effect of merging on halo and SMBH evolution.

Although our model considers only the effect of mergers of SMBHs and thus gives a lower limit on SMBH mass, it still leaves plenty of room to consider and include gas accretion. If the mass of SMBH at z=0 indicated from our model is found to be lower than observed, then the remaining mass should be explained by gas accretion. Since, in our model, the mass relation at any redshift other than $z=0$ is a free parameter and we provide a lot of freedom to pick the mass of initial black holes at $z=6$, an extra mass can be added to the black holes at each step in our merger tree, such that the SMBH mass at $z=0$ is conserved, as the SMBH mass at $z=6$ can be tuned accordingly using the amplitude and the power-law index. This way different accretion models can be introduced to our model.
Moreover, by fixing parameters like limiting mass and initial powerlaw index at $z=6$, our model can not only describe the observations at both $z=0$ and $z=6$ for the mass relationship between SMBH and DM halo, but also provide insight into the effect of accretion on the mass growth of SMBHs. 
Fixing these parameters is a major challenge because of the observational limitations. However, with better detectors, including but not limited to James Webb Space Telescope (JWST) \citep{doi:10.1063/1.3518858,2021jwst.prop.2057S}, and gravitational wave detections from Laser Interferometer Space Antenna (LISA) in the near future \citep{2001PhRvD..65b2004C,babak2008resolving,10.1111/j.1365-2966.2005.08987.x}, this issue could be well resolved. 
JWST can provide more precise observations and evaluations for halo properties (e.g., mass, velocity dispersion and rotation), reducing the observational bias for high redshift halos~\citep{10.1093/mnras/staa324}. A good estimate for constraining the limiting mass could be provided from the black hole occupation fraction~\citep{2019ApJ...874..117B}. The authors of the paper assume that halos that have grown to be above a critical mass $M_c$ have a chance probability p of having produced a seed black hole by this point. Current study uses the same approach with p fixed at unity. Thus, the critical mass resembles the limiting mass.

Our model is also self-consistent because it accounts for the PSMF and halos at every step following this mass function.
However, a few disadvantages were noted regarding this method. 
It ignores many physical processes (e.g., gravitational dynamics and accretion onto the BHs) because the model is simple. 
Although the gravitational dynamics should not change the halo mass distribution much on time scales of the order of $z=6$ as our model is self-consistent, it may still affect the mass of the BHs. 

Even with the limitations, the approach can have a significant contribution to explaining the evolution of SMBHs inside dark matter halos. 
Studies on gravitational waves from SMBH mergers could provide great insight into this work because the gravitational wave's amplitude is based on the merger history~\citep{2004ApJ...615...19E}. 
The event rate is calculated using the halo merger rates from the extended Press-Schechter theory~\citep{2006MNRAS.371.1992E, 2001ApJ...558..535M, 2003ApJ...590..691W, 2005ApJ...623...23S,2005MNRAS.361.1145R} and cosmological N-body simulations~\citep{2008arXiv0805.3154M}.  Our model could also be used to evaluate SMBH merger rates and calculate gravitational wave background due to the large range of redshifts covered.
LISA, which is scheduled for launch in this coming decade, will use detectors that may be ideal frequencies to measure gravitational waves generated from SMBH mergers \citep{2020JCAP...11..055P}.

\section*{Data Availability}
The data underlying this article will be shared on reasonable request to the corresponding author. 

\section*{Acknowledgements}
We thank K.~Shimasaku for providing us with the compiled data of the halo-SMBH mass relation.
This work is supported in part by the JSPS grant numbers 18K03616, 17H01110, 21H04467, 21H05459, 21KK0050, 21K03533, 17H04830, 21H04494 and JST AIP Acceleration Research Grant JP20317829 and JST FOREST Program JPMJFR20352935.


\bibliography{main}{}

\begin{thebibliography}{}
\expandafter\ifx\csname natexlab\endcsname\relax\def\natexlab#1{#1}\fi
\providecommand{\url}[1]{\href{#1}{#1}}
\providecommand{\dodoi}[1]{doi:~\href{http://doi.org/#1}{\nolinkurl{#1}}}
\providecommand{\doeprint}[1]{\href{http://ascl.net/#1}{\nolinkurl{http://ascl.net/#1}}}
\providecommand{\doarXiv}[1]{\href{https://arxiv.org/abs/#1}{\nolinkurl{https://arxiv.org/abs/#1}}}

\bibitem[{{Appenzeller} \& {Fricke}(1972)}]{1972A&A....18...10A}
{Appenzeller}, I., \& {Fricke}, K. 1972, aap, 18, 10

\bibitem[{{Ba{\~n}ados} {et~al.}(2016){Ba{\~n}ados}, {Venemans}, {Decarli},
  {Farina}, {Mazzucchelli}, {Walter}, {Fan}, {Stern}, {Schlafly}, {Chambers},
  {Rix}, {Jiang}, {McGreer}, {Simcoe}, {Wang}, {Yang}, {Morganson}, {De Rosa},
  {Greiner}, {Balokovi{\'c}}, {Burgett}, {Cooper}, {Draper}, {Flewelling},
  {Hodapp}, {Jun}, {Kaiser}, {Kudritzki}, {Magnier}, {Metcalfe}, {Miller},
  {Schindler}, {Tonry}, {Wainscoat}, {Waters}, \& {Yang}}]{2016ApJS..227...11B}
{Ba{\~n}ados}, E., {Venemans}, B.~P., {Decarli}, R., {et~al.} 2016, apjs, 227,
  11, \dodoi{10.3847/0067-0049/227/1/11}

\bibitem[{{Ba{\~n}ados} {et~al.}(2018){Ba{\~n}ados}, {Venemans},
  {Mazzucchelli}, {Farina}, {Walter}, {Wang}, {Decarli}, {Stern}, {Fan},
  {Davies}, {Hennawi}, {Simcoe}, {Turner}, {Rix}, {Yang}, {Kelson}, {Rudie}, \&
  {Winters}}]{2018Natur.553..473B}
{Ba{\~n}ados}, E., {Venemans}, B.~P., {Mazzucchelli}, C., {et~al.} 2018, nat,
  553, 473, \dodoi{10.1038/nature25180}

\bibitem[{Babak {et~al.}(2008)Babak, Hannam, Husa, \&
  Schutz}]{babak2008resolving}
Babak, S., Hannam, M., Husa, S., \& Schutz, B. 2008, Resolving Super Massive
  Black Holes with LISA.
\newblock \doarXiv{0806.1591}

\bibitem[{{Bandara} {et~al.}(2009){Bandara}, {Crampton}, \&
  {Simard}}]{2009ApJ...704.1135B}
{Bandara}, K., {Crampton}, D., \& {Simard}, L. 2009, apj, 704, 1135,
  \dodoi{10.1088/0004-637X/704/2/1135}

\bibitem[{{Banik} {et~al.}(2019){Banik}, {Tan}, \&
  {Monaco}}]{2019MNRAS.483.3592B}
{Banik}, N., {Tan}, J.~C., \& {Monaco}, P. 2019, mnras, 483, 3592,
  \dodoi{10.1093/mnras/sty3298}

\bibitem[{{Baugh}(2006)}]{2006RPPh...69.3101B}
{Baugh}, C.~M. 2006, Reports on Progress in Physics, 69, 3101,
  \dodoi{10.1088/0034-4885/69/12/R02}

\bibitem[{{Bond} {et~al.}(1991){Bond}, {Cole}, {Efstathiou}, \&
  {Kaiser}}]{1991ApJ...379..440B}
{Bond}, J.~R., {Cole}, S., {Efstathiou}, G., \& {Kaiser}, N. 1991, apj, 379,
  440, \dodoi{10.1086/170520}

\bibitem[{Bower {et~al.}(2016)Bower, Schaye, Frenk, Theuns, Schaller, Crain, \&
  McAlpine}]{10.1093/mnras/stw2735}
Bower, R.~G., Schaye, J., Frenk, C.~S., {et~al.} 2016, Monthly Notices of the
  Royal Astronomical Society, 465, 32, \dodoi{10.1093/mnras/stw2735}

\bibitem[{{Bromm} \& {Loeb}(2003)}]{2003ApJ...596...34B}
{Bromm}, V., \& {Loeb}, A. 2003, apj, 596, 34, \dodoi{10.1086/377529}

\bibitem[{{Buchner} {et~al.}(2019){Buchner}, {Treister}, {Bauer}, {Sartori}, \&
  {Schawinski}}]{2019ApJ...874..117B}
{Buchner}, J., {Treister}, E., {Bauer}, F.~E., {Sartori}, L.~F., \&
  {Schawinski}, K. 2019, apj, 874, 117, \dodoi{10.3847/1538-4357/aafd32}

\bibitem[{{Chon} \& {Omukai}(2020)}]{2020MNRAS.494.2851C}
{Chon}, S., \& {Omukai}, K. 2020, mnras, 494, 2851,
  \dodoi{10.1093/mnras/staa863}

\bibitem[{{Cornish}(2001)}]{2001PhRvD..65b2004C}
{Cornish}, N.~J. 2001, \prd, 65, 022004, \dodoi{10.1103/PhysRevD.65.022004}

\bibitem[{{Davies} {et~al.}(2011){Davies}, {Miller}, \&
  {Bellovary}}]{2011ApJ...740L..42D}
{Davies}, M.~B., {Miller}, M.~C., \& {Bellovary}, J.~M. 2011, apjl, 740, L42,
  \dodoi{10.1088/2041-8205/740/2/L42}

\bibitem[{{Davis} {et~al.}(2019){Davis}, {Graham}, \&
  {Combes}}]{2019ApJ...877...64D}
{Davis}, B.~L., {Graham}, A.~W., \& {Combes}, F. 2019, apj, 877, 64,
  \dodoi{10.3847/1538-4357/ab1aa4}

\bibitem[{de Nicola {et~al.}(2019)de Nicola, Marconi, \&
  Longo}]{10.1093/mnras/stz2472}
de Nicola, S., Marconi, A., \& Longo, G. 2019, Monthly Notices of the Royal
  Astronomical Society, 490, 600, \dodoi{10.1093/mnras/stz2472}

\bibitem[{{Djorgovski} {et~al.}(2008){Djorgovski}, {Volonteri}, {Springel},
  {Bromm}, \& {Meylan}}]{2008mgm..conf..340D}
{Djorgovski}, S.~G., {Volonteri}, M., {Springel}, V., {Bromm}, V., \& {Meylan},
  G. 2008, in The Eleventh Marcel Grossmann Meeting On Recent Developments in
  Theoretical and Experimental General Relativity, Gravitation and Relativistic
  Field Theories, 340--367, \dodoi{10.1142/9789812834300\_0018}

\bibitem[{Endsley {et~al.}(2020)Endsley, Behroozi, Stark, Williams, Robertson,
  Rieke, Gottlöber, \& Yepes}]{10.1093/mnras/staa324}
Endsley, R., Behroozi, P., Stark, D.~P., {et~al.} 2020, Monthly Notices of the
  Royal Astronomical Society, 493, 1178, \dodoi{10.1093/mnras/staa324}

\bibitem[{{Enoki} {et~al.}(2004){Enoki}, {Inoue}, {Nagashima}, \&
  {Sugiyama}}]{2004ApJ...615...19E}
{Enoki}, M., {Inoue}, K.~T., {Nagashima}, M., \& {Sugiyama}, N. 2004, apj, 615,
  19, \dodoi{10.1086/424475}

\bibitem[{{Erickcek} {et~al.}(2006){Erickcek}, {Kamionkowski}, \&
  {Benson}}]{2006MNRAS.371.1992E}
{Erickcek}, A.~L., {Kamionkowski}, M., \& {Benson}, A.~J. 2006, mnras, 371,
  1992, \dodoi{10.1111/j.1365-2966.2006.10838.x}

\bibitem[{{Fan} {et~al.}(2001){Fan}, {Narayanan}, {Lupton}, {Strauss}, {Knapp},
  {Becker}, {White}, {Pentericci}, {Leggett}, {Haiman}, {Gunn}, {Ivezi{\'c}},
  {Schneider}, {Anderson}, {Brinkmann}, {Bahcall}, {Connolly}, {Csabai}, {Doi},
  {Fukugita}, {Geballe}, {Grebel}, {Harbeck}, {Hennessy}, {Lamb}, {Miknaitis},
  {Munn}, {Nichol}, {Okamura}, {Pier}, {Prada}, {Richards}, {Szalay}, \&
  {York}}]{2001AJ....122.2833F}
{Fan}, X., {Narayanan}, V.~K., {Lupton}, R.~H., {et~al.} 2001, aj, 122, 2833,
  \dodoi{10.1086/324111}

\bibitem[{{Ferrarese}(2002)}]{2002ApJ...578...90F}
{Ferrarese}, L. 2002, apj, 578, 90,
  \dodoi{10.1086/34230810.48550/arXiv.astro-ph/0203469}

\bibitem[{{Ferrarese} \& {Merritt}(2000)}]{2000ApJ...539L...9F}
{Ferrarese}, L., \& {Merritt}, D. 2000, apjl, 539, L9, \dodoi{10.1086/312838}

\bibitem[{{Griffin} {et~al.}(2019){Griffin}, {Lacey}, {Gonzalez-Perez},
  {Lagos}, {Baugh}, \& {Fanidakis}}]{2019MNRAS.487..198G}
{Griffin}, A.~J., {Lacey}, C.~G., {Gonzalez-Perez}, V., {et~al.} 2019, mnras,
  487, 198, \dodoi{10.1093/mnras/stz1216}

\bibitem[{{Guo} {et~al.}(2013){Guo}, {White}, {Angulo}, {Henriques}, {Lemson},
  {Boylan-Kolchin}, {Thomas}, \& {Short}}]{2013MNRAS.428.1351G}
{Guo}, Q., {White}, S., {Angulo}, R.~E., {et~al.} 2013, mnras, 428, 1351,
  \dodoi{10.1093/mnras/sts115}

\bibitem[{{Haemmerl{\'e}}(2021)}]{2021A&A...650A.204H}
{Haemmerl{\'e}}, L. 2021, aap, 650, A204, \dodoi{10.1051/0004-6361/202140893}

\bibitem[{{Kauffmann} \& {Haehnelt}(2000)}]{2000MNRAS.311..576K}
{Kauffmann}, G., \& {Haehnelt}, M. 2000, mnras, 311, 576,
  \dodoi{10.1046/j.1365-8711.2000.03077.x}

\bibitem[{{King} {et~al.}(2008){King}, {Pringle}, \&
  {Hofmann}}]{2008MNRAS.385.1621K}
{King}, A.~R., {Pringle}, J.~E., \& {Hofmann}, J.~A. 2008, mnras, 385, 1621,
  \dodoi{10.1111/j.1365-2966.2008.12943.x10.48550/arXiv.0801.1564}

\bibitem[{{Kormendy} \& {Ho}(2013)}]{2013ARA&A..51..511K}
{Kormendy}, J., \& {Ho}, L.~C. 2013, araa, 51, 511,
  \dodoi{10.1146/annurev-astro-082708-101811}

\bibitem[{{Kroupa} {et~al.}(2020){Kroupa}, {Subr}, {Jerabkova}, \&
  {Wang}}]{2020MNRAS.498.5652K}
{Kroupa}, P., {Subr}, L., {Jerabkova}, T., \& {Wang}, L. 2020, mnras, 498,
  5652, \dodoi{10.1093/mnras/staa2276}

\bibitem[{Kroupa {et~al.}(2020)Kroupa, Subr, Jerabkova, \&
  Wang}]{10.1093/mnras/staa2276}
Kroupa, P., Subr, L., Jerabkova, T., \& Wang, L. 2020, Monthly Notices of the
  Royal Astronomical Society, 498, 5652, \dodoi{10.1093/mnras/staa2276}

\bibitem[{{Krumpe} {et~al.}(2015){Krumpe}, {Miyaji}, {Husemann}, {Fanidakis},
  {Coil}, \& {Aceves}}]{2015ApJ...815...21K}
{Krumpe}, M., {Miyaji}, T., {Husemann}, B., {et~al.} 2015, apj, 815, 21,
  \dodoi{10.1088/0004-637X/815/1/21}

\bibitem[{{Lacey} \& {Cole}(1993)}]{1993MNRAS.262..627L}
{Lacey}, C., \& {Cole}, S. 1993, mnras, 262, 627,
  \dodoi{10.1093/mnras/262.3.627}

\bibitem[{{Malbon} {et~al.}(2007){Malbon}, {Baugh}, {Frenk}, \&
  {Lacey}}]{2007MNRAS.382.1394M}
{Malbon}, R.~K., {Baugh}, C.~M., {Frenk}, C.~S., \& {Lacey}, C.~G. 2007, mnras,
  382, 1394, \dodoi{10.1111/j.1365-2966.2007.12317.x}

\bibitem[{{Marasco} {et~al.}(2021){Marasco}, {Cresci}, {Posti}, {Fraternali},
  {Mannucci}, {Marconi}, {Belfiore}, \& {Fall}}]{2021MNRAS.507.4274M}
{Marasco}, A., {Cresci}, G., {Posti}, L., {et~al.} 2021, mnras, 507, 4274,
  \dodoi{10.1093/mnras/stab2317}

\bibitem[{{Matsuoka} {et~al.}(2019){Matsuoka}, {Iwasawa}, {Onoue}, {Kashikawa},
  {Strauss}, {Lee}, {Imanishi}, {Nagao}, {Akiyama}, {Asami}, {Bosch},
  {Furusawa}, {Goto}, {Gunn}, {Harikane}, {Ikeda}, {Izumi}, {Kawaguchi},
  {Kato}, {Kikuta}, {Kohno}, {Komiyama}, {Koyama}, {Lupton}, {Minezaki},
  {Miyazaki}, {Murayama}, {Niida}, {Nishizawa}, {Noboriguchi}, {Oguri}, {Ono},
  {Ouchi}, {Price}, {Sameshima}, {Schulze}, {Silverman}, {Sugiyama}, {Tait},
  {Takada}, {Takata}, {Tanaka}, {Tang}, {Toba}, {Utsumi}, {Wang}, \&
  {Yamashita}}]{2019ApJ...883..183M}
{Matsuoka}, Y., {Iwasawa}, K., {Onoue}, M., {et~al.} 2019, apj, 883, 183,
  \dodoi{10.3847/1538-4357/ab3c60}

\bibitem[{{Matsuoka} {et~al.}(2022){Matsuoka}, {Iwasawa}, {Onoue}, {Izumi},
  {Kashikawa}, {Strauss}, {Imanishi}, {Nagao}, {Akiyama}, {Silverman}, {Asami},
  {Bosch}, {Furusawa}, {Goto}, {Gunn}, {Harikane}, {Ikeda}, {Ishimoto},
  {Kawaguchi}, {Kato}, {Kikuta}, {Kohno}, {Komiyama}, {Lee}, {Lupton},
  {Minezaki}, {Miyazaki}, {Murayama}, {Nishizawa}, {Oguri}, {Ono}, {Ouchi},
  {Price}, {Sameshima}, {Sugiyama}, {Tait}, {Takada}, {Takahashi}, {Takata},
  {Tanaka}, {Toba}, {Utsumi}, {Wang}, \& {Yamashita}}]{2022ApJS..259...18M}
---. 2022, apjs, 259, 18, \dodoi{10.3847/1538-4365/ac3d31}

\bibitem[{{Menou} {et~al.}(2001){Menou}, {Haiman}, \&
  {Narayanan}}]{2001ApJ...558..535M}
{Menou}, K., {Haiman}, Z., \& {Narayanan}, V.~K. 2001, apj, 558, 535,
  \dodoi{10.1086/322310}

\bibitem[{{Micic} {et~al.}(2008){Micic}, {Holley-Bockelmann}, \&
  {Sigurdsson}}]{2008arXiv0805.3154M}
{Micic}, M., {Holley-Bockelmann}, K., \& {Sigurdsson}, S. 2008, arXiv e-prints,
  arXiv:0805.3154.
\newblock \doarXiv{0805.3154}

\bibitem[{{Mo} {et~al.}(2010){Mo}, {van den Bosch}, \&
  {White}}]{2010gfe..book.....M}
{Mo}, H., {van den Bosch}, F.~C., \& {White}, S. 2010, {Galaxy Formation and
  Evolution}

\bibitem[{{Padmanabhan} \& {Loeb}(2020)}]{2020JCAP...11..055P}
{Padmanabhan}, H., \& {Loeb}, A. 2020, \jcap, 2020, 055,
  \dodoi{10.1088/1475-7516/2020/11/055}

\bibitem[{{Portegies Zwart} \& {McMillan}(2002)}]{2002ApJ...576..899P}
{Portegies Zwart}, S.~F., \& {McMillan}, S. L.~W. 2002, apj, 576, 899,
  \dodoi{10.1086/341798}

\bibitem[{{Press} \& {Schechter}(1974)}]{1974ApJ...187..425P}
{Press}, W.~H., \& {Schechter}, P. 1974, apj, 187, 425, \dodoi{10.1086/152650}

\bibitem[{Rhook \& Wyithe(2005)}]{10.1111/j.1365-2966.2005.08987.x}
Rhook, K.~J., \& Wyithe, J. S.~B. 2005, Monthly Notices of the Royal
  Astronomical Society, 361, 1145, \dodoi{10.1111/j.1365-2966.2005.08987.x}

\bibitem[{{Rhook} \& {Wyithe}(2005)}]{2005MNRAS.361.1145R}
{Rhook}, K.~J., \& {Wyithe}, J. S.~B. 2005, mnras, 361, 1145,
  \dodoi{10.1111/j.1365-2966.2005.08987.x}

\bibitem[{{Richstone} {et~al.}(1998){Richstone}, {Ajhar}, {Bender}, {Bower},
  {Dressler}, {Faber}, {Filippenko}, {Gebhardt}, {Green}, {Ho}, {Kormendy},
  {Lauer}, {Magorrian}, \& {Tremaine}}]{1998Natur.395A..14R}
{Richstone}, D., {Ajhar}, E.~A., {Bender}, R., {et~al.} 1998, nat, 385, A14.
\newblock \doarXiv{astro-ph/9810378}

\bibitem[{{Rossi} {et~al.}(2010){Rossi}, {Lodato}, {Armitage}, {Pringle}, \&
  {King}}]{2010MNRAS.401.2021R}
{Rossi}, E.~M., {Lodato}, G., {Armitage}, P.~J., {Pringle}, J.~E., \& {King},
  A.~R. 2010, mnras, 401, 2021, \dodoi{10.1111/j.1365-2966.2009.15802.x}

\bibitem[{{Sabra} {et~al.}(2008){Sabra}, {Akl}, \&
  {Chahine}}]{2008IAUS..245..257S}
{Sabra}, B.~M., {Akl}, M.~A., \& {Chahine}, G. 2008, in Formation and Evolution
  of Galaxy Bulges, ed. M.~{Bureau}, E.~{Athanassoula}, \& B.~{Barbuy}, Vol.
  245, 257--258, \dodoi{10.1017/S1743921308017857}

\bibitem[{{Sesana} {et~al.}(2005){Sesana}, {Haardt}, {Madau}, \&
  {Volonteri}}]{2005ApJ...623...23S}
{Sesana}, A., {Haardt}, F., {Madau}, P., \& {Volonteri}, M. 2005, apj, 623, 23,
  \dodoi{10.1086/428492}

\bibitem[{{Shen} {et~al.}(2021){Shen}, {Li}, {Anderson}, {Brandt}, {Hall},
  {Ho}, {Homayouni}, {Horne}, {Ibarra Medel}, {Jiang}, {Liu}, {Peterson},
  {Schneider}, {Trump}, \& {Yang}}]{2021jwst.prop.2057S}
{Shen}, Y., {Li}, J.~I., {Anderson}, S.~F., {et~al.} 2021, {A JWST Study of the
  Link Between Supermassive Black Holes and Galaxies at Cosmic Noon}, JWST
  Proposal. Cycle 1

\bibitem[{{Shimasaku} \& {Izumi}(2019)}]{2019ApJ...872L..29S}
{Shimasaku}, K., \& {Izumi}, T. 2019, apjl, 872, L29,
  \dodoi{10.3847/2041-8213/ab053f}

\bibitem[{{Somerville} \& {Kolatt}(1999)}]{1999MNRAS.305....1S}
{Somerville}, R.~S., \& {Kolatt}, T.~S. 1999, mnras, 305, 1,
  \dodoi{10.1046/j.1365-8711.1999.02154.x}

\bibitem[{{Venemans} {et~al.}(2017){Venemans}, {Walter}, {Decarli},
  {Ba{\~n}ados}, {Carilli}, {Winters}, {Schuster}, {da Cunha}, {Fan}, {Farina},
  {Mazzucchelli}, {Rix}, \& {Weiss}}]{2017ApJ...851L...8V}
{Venemans}, B.~P., {Walter}, F., {Decarli}, R., {et~al.} 2017, apjl, 851, L8,
  \dodoi{10.3847/2041-8213/aa943a}

\bibitem[{{Vergara} {et~al.}(2021){Vergara}, {Schleicher}, {Boekholt},
  {Reinoso}, {Fellhauer}, {Klessen}, \& {Leigh}}]{2021A&A...649A.160V}
{Vergara}, M.~Z.~C., {Schleicher}, D.~R.~G., {Boekholt}, T.~C.~N., {et~al.}
  2021, aap, 649, A160, \dodoi{10.1051/0004-6361/202140298}

\bibitem[{{Volonteri}(2010)}]{2010A&ARv..18..279V}
{Volonteri}, M. 2010, aapr, 18, 279, \dodoi{10.1007/s00159-010-0029-x}

\bibitem[{{Wang} {et~al.}(2021){Wang}, {Yang}, {Fan}, {Hennawi}, {Barth},
  {Banados}, {Bian}, {Boutsia}, {Connor}, {Davies}, {Decarli}, {Eilers},
  {Farina}, {Green}, {Jiang}, {Li}, {Mazzucchelli}, {Nanni}, {Schindler},
  {Venemans}, {Walter}, {Wu}, \& {Yue}}]{2021ApJ...907L...1W}
{Wang}, F., {Yang}, J., {Fan}, X., {et~al.} 2021, apjl, 907, L1,
  \dodoi{10.3847/2041-8213/abd8c6}

\bibitem[{{Whitmore} {et~al.}(1979){Whitmore}, {Kirshner}, \&
  {Schechter}}]{1979ApJ...234...68W}
{Whitmore}, B.~C., {Kirshner}, R.~P., \& {Schechter}, P.~L. 1979, apj, 234, 68,
  \dodoi{10.1086/157473}

\bibitem[{{Willott} {et~al.}(2010{\natexlab{a}}){Willott}, {Delorme},
  {Reyl{\'e}}, {Albert}, {Bergeron}, {Crampton}, {Delfosse}, {Forveille},
  {Hutchings}, {McLure}, {Omont}, \& {Schade}}]{2010AJ....139..906W}
{Willott}, C.~J., {Delorme}, P., {Reyl{\'e}}, C., {et~al.} 2010{\natexlab{a}},
  aj, 139, 906, \dodoi{10.1088/0004-6256/139/3/906}

\bibitem[{{Willott} {et~al.}(2010{\natexlab{b}}){Willott}, {Albert},
  {Arzoumanian}, {Bergeron}, {Crampton}, {Delorme}, {Hutchings}, {Omont},
  {Reyl{\'e}}, \& {Schade}}]{2010AJ....140..546W}
{Willott}, C.~J., {Albert}, L., {Arzoumanian}, D., {et~al.} 2010{\natexlab{b}},
  aj, 140, 546, \dodoi{10.1088/0004-6256/140/2/546}

\bibitem[{Windhorst \& Cohen(2010)}]{doi:10.1063/1.3518858}
Windhorst, R.~A., \& Cohen, S.~H. 2010, AIP Conference Proceedings, 1294, 225,
  \dodoi{10.1063/1.3518858}

\bibitem[{{Wyithe} \& {Loeb}(2003)}]{2003ApJ...590..691W}
{Wyithe}, J. S.~B., \& {Loeb}, A. 2003, apj, 590, 691, \dodoi{10.1086/375187}

\bibitem[{{Yang} {et~al.}(2020){Yang}, {Wang}, {Fan}, {Hennawi}, {Davies},
  {Yue}, {Banados}, {Wu}, {Venemans}, {Barth}, {Bian}, {Boutsia}, {Decarli},
  {Farina}, {Green}, {Jiang}, {Li}, {Mazzucchelli}, \&
  {Walter}}]{2020ApJ...897L..14Y}
{Yang}, J., {Wang}, F., {Fan}, X., {et~al.} 2020, apjl, 897, L14,
  \dodoi{10.3847/2041-8213/ab9c26}

\bibitem[{{Yu} \& {Tremaine}(2002)}]{2002MNRAS.335..965Y}
{Yu}, Q., \& {Tremaine}, S. 2002, mnras, 335, 965,
  \dodoi{10.1046/j.1365-8711.2002.05532.x}

\bibitem[{{Zhang} {et~al.}(2021){Zhang}, {Guo}, {Qu}, \&
  {Gao}}]{2021RAA....21..212Z}
{Zhang}, T.-C., {Guo}, Q., {Qu}, Y., \& {Gao}, L. 2021, Research in Astronomy
  and Astrophysics, 21, 212, \dodoi{10.1088/1674-4527/21/8/212}

\bibitem[{{Zubovas} \& {King}(2021)}]{2021MNRAS.501.4289Z}
{Zubovas}, K., \& {King}, A. 2021, mnras, 501, 4289,
  \dodoi{10.1093/mnras/stab004}

\end{thebibliography}
\bibliographystyle{aasjournal}

\end{document}